\documentclass{article}
\usepackage{graphicx} % Required for inserting images
\usepackage[caption=false]{subfig}
\usepackage{hyperref}

\title{Privacy-Preserving Decentralized AI with Confidential Computing}
\author{Dayeol Lee, Jorge Ant\'onio, Hisham Khan}
\date{October 2024}

\begin{document}

\maketitle

\section{Introduction}

In this paper, we discuss privacy protection based on Confidential Computing (CC) in Atoma Network, which aims to decentralize Artificial Intelligence (AI).
Decentralized AI aims to distribute AI inference or training services between multiple entities so that there is no central authority.
We believe that decentralization is a crucial technology for trustworthy, transparent, and robust AI.
We truly believe that Atoma, a decentralized network tailored to AI, will unlock a lot of AI applications in the Web3 domain.

One of the primary challenges in the realm of decentralized AI is safeguarding privacy. Decentralized AI enables a diverse ecosystem where node operators, model providers, data providers, and end-users can all contribute and interact within the network without the need for central oversight or permissions. This openness promotes innovation and broader access, but it also introduces significant privacy concerns.

Consider, for instance, data centers equipped with AI accelerators such as GPUs. These centers can contribute their computing power to the network, allowing for distributed processing and scaling. Simultaneously, owners of large foundational models, whether proprietary or open source, can deploy these models to the decentralized system, making them available for various use cases. End-users, on the other hand, can run AI inference or training workloads using their own personal data, benefiting from the network's distributed capabilities.

However, this decentralized structure inherently implies that sensitive assets, such as proprietary models, confidential data, and personal information, may be shared among a network of participants who may not always be trustworthy. The permissionless nature of such systems means that not all participants are subject to stringent regulations or bound by reputational concerns, thus creating an environment where privacy violations could  occur easily. Incentives to protect privacy might be misaligned, as some participants may prioritize financial gain over ethical considerations.

Furthermore, the decentralized nature of the network complicates the enforcement of privacy standards. Traditional systems rely on centralized control to enforce privacy policies, regulate access, and ensure accountability. However, in a decentralized system, the lack of a singular governing body makes it difficult to ensure that privacy is consistently respected. Participants may engage in data leaks, model theft, or unauthorized access to proprietary information without fear of legal consequences or damage to their reputation.

% [@jorge] That said, different techniques have been explored to provide trust for AI compute. Some of these techniques include zero-knowledge machine learning (zkML), optimistic machine learning (opML), or even our internally developed Sampling Consensus techniques. That said, zkML while providing model weight privacy (by running models on owned hardware and providing proofs of the execution correctness) is currently far from being mature enough to be applied to Large Language Models (LLMs) computed. In particular, zkML might impose many orders of magnitude of compute and latency overhead for processing LLM requests (see \cite{}). On the other hand, while the Sampling Consensus and opML protocols have been applied in production, recently, to provide high guarantees of the correctness of AI generated outputs, these techniques do not offer any means of data or model weight privacy.

Various techniques are being explored to bolster trust in AI computation, with methods such as zero-knowledge machine learning (zkML), optimistic machine learning (opML), and Atoma's innovative Sampling Consensus protocol standing out. zkML, for instance, offers model weight privacy by allowing machine learning models to be executed on owned hardware while generating cryptographic proofs that verify the correctness of the computation. This ensures that model weights remain private, shielding proprietary information from other network participants. However, zkML, in its current state, is not yet mature enough for the demands of Large Language Model (LLM) applications. The computational and latency overhead associated with zkML is prohibitive for LLMs, often increasing resource requirements by several orders of magnitude, as demonstrated in prior work (see \cite{sun2024zkllmzeroknowledgeproofs}). Unfortunately, zkML offers no data privacy for both end users and developers. Any data to be processed by a model needs to be sent in plain-text to the node operator, so that the latter can generate the appropriate zk-proofs.

On the other hand, Sampling Consensus and opML are a secure, cheaper alternative to zkML, and production-ready for ensuring the integrity and correctness of AI-generated outputs through consensus and optimistic verification mechanisms. These techniques excel at verifying output correctness in distributed networks, making them practical for real-world decentralized AI applications. However, despite their success in ensuring computational accuracy, they fall short of addressing a critical concern: the privacy of data and model weight parameters. Neither protocol directly tackles the challenge of protecting sensitive information, leaving privacy concerns unresolved in many decentralized AI deployments.

In this paper, we explore the potential of confidential computing (CC) as a promising solution to bridge this privacy gap. Confidential computing leverages hardware-based Trusted Execution Environments (TEEs) to protect the integrity of code and the confidentiality of data processed on remote machines. TEEs provide an isolated, secure environment where sensitive computations can take place, ensuring that neither the data nor the model parameters are exposed to other participants in the network, even in potentially untrusted environments. 
In the context of Atoma, TEEs can ensure that proprietary models and personal data remain confidential, even when run on distributed nodes. This hardware-enforced security model offers a practical path to addressing the privacy concerns that currently limit the adoption of decentralized AI solutions.

Admittedly, TEEs are still an early-stage technology and come with their own set of limitations. These include challenges related to potential hardware vulnerabilities. Despite these challenges, we argue that TEEs represent a significant step forward in bringing privacy into decentralized AI. Their ability to offer strong privacy protections, combined with their hardware-based nature, makes them a promising candidate for secure and scalable decentralized AI infrastructure. In particular, we will examine how TEEs can enhance privacy within Atoma's decentralized framework, offering a path to more secure and trustworthy AI computations in the future.

\section{Privacy in Decentralized AI}

Protecting privacy is fundamental to the success of decentralized AI. In a decentralized environment, permissionless participation allows anyone to join the network as a node operator, providing computational power in exchange for rewards. However, this openness introduces significant risks, as sensitive data—including inputs, outputs, model weights, and intermediate computational states—can reside in the memory of nodes operated by random, and potentially untrustworthy, participants. Additionally, some of these values may need to be publicly visible on-chain for purposes of transparency and verifiability, further complicating the privacy landscape.

Unfortunately, many aspects of AI workloads require confidentiality and, without proper privacy protections, the decentralized AI ecosystem may struggle to gain the trust of key stakeholders. To illustrate this, consider the following scenarios that demonstrate why privacy is crucial in decentralized AI systems.

\begin{itemize}
\item\textbf{Inference.} Many AI inference use cases involve personal or proprietary information. For example, users running AI-powered health diagnostics or business-specific predictive models may want to keep both the inputs (e.g., personal data) and the outputs (e.g., predictions) private. Without privacy, users would be reluctant to take advantage of decentralized AI for sensitive applications, stifling its potential.

\item\textbf{Model Serving.}
Owners of foundational models can seek to monetize their proprietary models by deploying them on a decentralized network. However, without strong privacy guarantees, the risk of model theft increases significantly, as malicious actors could extract or replicate the model weights. This would disincentivize private model owners from contributing their valuable resources to the network, limiting the variety and richness of models available in the decentralized ecosystem.

\item\textbf{Intellectual Property.} Developers who create application-specific models or prompts rely on protecting their intellectual property (IP). In a completely open, decentralized network without privacy safeguards, sensitive IP can be exposed, especially in jurisdictions with weaker regulations. The lack of data privacy mechanisms could result in widespread leaks, diminishing the value of proprietary technologies, and discouraging the development of sophisticated applications on decentralized AI platforms.

\item\textbf{Model Training.}
Decentralized AI offers the exciting possibility of collaborative model training, where participants can contribute datasets to enhance model performance. However, data owners are unlikely to participate if they are required to directly expose their data to others. Many would prefer privacy-preserving techniques such as differential privacy, where the data is anonymized and protected, or encrypted data that is used only for computation and then discarded. Without such protections, data owners will be reluctant to share their valuable datasets, hindering the collaborative potential of decentralized AI.

\end{itemize}

We believe that decentralized AI will be one of the key pillars in future AI technology, unlocking many applications that are not really possible in AI services governed by a single entity.
However, lack of privacy would limit the true power of decentralized AI.

\section{Confidential Computing for Decentralized AI}

Confidential Computing is a promising technology that can protect applications from the rest of the system by using a hardware-based isolation technique called trusted execution environment (TEE).
TEEs aim to isolate an application during its execution from any other software, even the operating system. 
This means that they could protect the confidentiality and integrity of the application while it's running on a remote machine that is owned and operated by an untrusted operator.
Thus, TEEs can offer privacy even in a decentralized setting, where anyone can participate to the network as a node operator.
 
Cryptographic solutions like zkML exist, but they are not yet equipped to handle AI workloads requiring extensive computation and large memory use.

To provide context, the zkLLM work cited in \cite{sun2024zkllmzeroknowledgeproofs} is currently seen as the leading zkML protocol that aims to achieve verifiable LLM inference. It utilizes GPU acceleration with the CUDA library in \cite{filecoin-ecgpu} to handle BLS12-381 curve operations on NVIDIA GPUs. Notably, zkLLM achieves a 15-minute proof generation time for a single 13 billion-parameter LLM inference forward pass, which predicts the next token in the sequence. However, producing an entire output sequence of 100 tokens (a typical short answer) requires aggregating 100 forward pass zk proofs, resulting in at least 1500 minutes (or 25 hours) for full LLM inference proof generation. Furthermore, the total proving time is expected to be significantly higher because the attention Key and Value (KV) caches grow linearly, while the softmax scores expand quadratically with sequence length. Given that GPU memory, rather than compute capabilities, constrains LLM inference on GPUs, expanding KV caches will likely lead to even greater proof generation times in practice.

In contrast, TEEs offer small overhead as it relies on hardware-based memory isolation and encryption, and allow processors to compute on plain text. They are readily available in most future-generation server processors by major vendors.
Intel has been including Trusted Domain Extension (TDX) since 5th generation Xeon server processors (Emerald Rapids) and onwards, and AMD has been including Secure Encrypted Virtualization for Secure Nested Paging (SEV-SNP) since the Milan generation of EPYC server processors and onwards.
Also, NVIDIA has been including Confidential Computing as its main security feature starting with the Hopper architecture.
Confidential computing capability allows running unmodified AI applications with very small performance overhead on the next-generation systems.
Thus, we believe that TEEs are the best way to fill the gap until the other techniques mature.

\section{Leveraging Confidential Computing for Privacy in Decentralized AI}

\subsection{Overview}
In this section, we explore how Atoma leverages Trusted Execution Environments (TEEs) within its execution nodes to enhance privacy and secure AI workloads. By utilizing TEEs, Atoma ensures that sensitive data, such as the inputs and outputs of AI tasks, remain protected from the node operators who host the execution environments. For a deeper discussion on model privacy, refer to Section~\ref{sec:future-model-privacy}.

To incentivize privacy-focused infrastructure, node operators are rewarded for offering TEE-capable hardware, with additional compensation for maintaining these capabilities. When network users request privacy for their AI workloads, Atoma intelligently routes their requests to execution nodes equipped with TEEs. TEEs ensure that inputs, outputs, and models are decrypted only within an isolated and secure environment—such as a virtual machine with a predefined initial state—ensuring that even the node operators, who control the underlying hardware, cannot access the sensitive data. This encryption process is further secured by provisioning decryption keys exclusively to the specific TEE instance handling the workload, thereby mitigating risks of unauthorized access.

\begin{figure}
    \centering
    \includegraphics[width=0.7\linewidth]{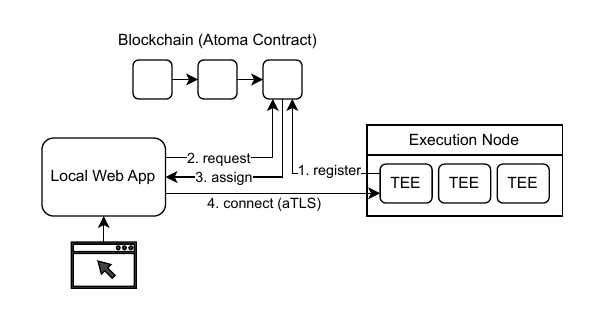}
    \caption{User Attestation}
    \label{fig:user-attestation}
\end{figure}

\begin{figure}
    \centering
    \includegraphics[width=0.7\linewidth]{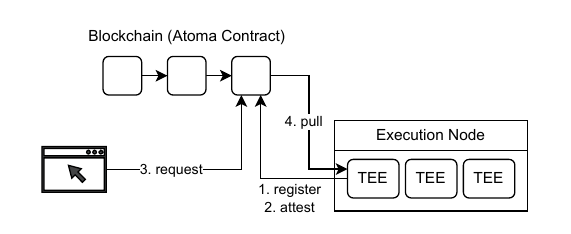}
    \caption{On-Chain Attestation}
    \label{fig:onchain-attestation}
\end{figure}

Atoma's privacy architecture includes two distinct approaches for key provisioning and attestation: User Attestation (Figure~\ref{fig:user-attestation}) and On-Chain Attestation (Figure~\ref{fig:onchain-attestation}). These approaches ensure that the integrity and privacy of the execution environment are verified before sensitive data is decrypted, adding an additional layer of trust to the system. Through this integration of TEEs and robust attestation mechanisms, Atoma offers a strong solution for maintaining privacy in decentralized AI networks.

% [@DLee] How is contract supposed to help the client to directly establish a session with the TEE ? The contract selects the node(s) to perform the computation, each node will have a unique identifier that is generated by the contract. We can have a middleware that can be queried to establish the connection between the user and node, if that makes sense.

% [@jorge] This is correct, we are exploring possible middleware layers to establish direct communication between users and nodes. We could use a p2p network like Waku for the node and the user negotiate establishing the connection. Once that is done, the user should be able to communicate with the node via HTTP requests (following OpenAI api), or other forms more suitable for other applications than inference.

\subsection{User Attestation}

\paragraph{Node Registration.} During the initialization phase, node operators are responsible for launching statically configured instances on their execution nodes. If their hardware—both CPU and GPU—supports Trusted Execution Environment (TEE) instances, operators can choose to deploy these secure environments. Depending on the memory and computational requirements of the models being executed, each node can run multiple TEE instances. Once a TEE instance is up and running, it will autonomously \textit{register} itself with the Atoma contract, signaling its readiness to handle privacy-sensitive workloads.

\paragraph{User Request.} Users can interact with the Atoma contract through a secure, open-source local web application. As we will explain in further detail, this local web app is essential for facilitating the Attested Transport Layer Security (aTLS) protocol, which provides a secure communication channel. Using their browser, the user can seamlessly interact with the local web app. When the user submits a request to the Atoma contract, the system dynamically assigns a specific TEE instance to handle the request, ensuring that the workload is processed securely.

\paragraph{Connect via aTLS.} Once the user is assigned a node, they establish a direct connection to the TEE instance through the aTLS protocol. This connection can be managed through middleware, which translates the unique identifier of the assigned TEE instance into the relevant metadata required for communication. The aTLS protocol binds the user's TLS session to the TEE’s attestation report, enabling the user to verify the integrity and authenticity of the TEE before any data exchange occurs. Upon successful verification of the attestation report, the user can confidently begin transmitting sensitive data through an encrypted channel. While attestation takes place at the start of each TLS session, the certificate can be cached on the client side to reduce overhead and streamline future connections, providing a balance between security and performance.

\subsubsection{Discussion}

For User Attestation, a local web application is required to facilitate the aTLS protocol, as current web browsers do not natively support attestation within the TLS handshake process. A proposal has been submitted to the IETF \cite{ietf-atls-proposal} to integrate attestation directly into the TLS 1.3 handshake, but widespread standardization and browser support remain distant prospects. As a result, the most practical solution for enabling this non-standard encrypted communication channel is to utilize a local web application, which acts as an intermediary for secure communication.

The aTLS protocol requires trust to be established at the beginning of each TLS session through attestation, ensuring that the TEE instance can reliably prove its integrity and operational state. This initial verification process effectively mitigates Time-of-Check to Time-of-Use (TOCTOU) attacks, as well as impersonation attempts, because the TEE must provide valid attestation proof before any sensitive data is exchanged. If a TEE instance fails to produce a valid attestation report, the user can simply notify the Atoma contract and switch to an alternate node, preserving the integrity of the system.

One potential issue with User Attestation is that the user could falsify the attestation verification result, bailing out of the transaction after the node has already begun serving the request. To prevent such exploitation, we propose a mechanism where the user must notify the contract with the verified attestation result before the node commences the actual computation. By introducing this step, the network ensures that computations only proceed after successful attestation, preventing misuse and ensuring a more secure and transparent process for both users and node operators.

\subsection{On-Chain Attestation}

As an alternative to User Attestation, via the aTLS protocol, Atoma can leverage on-chain attestations, which leverage on-chain verification mechanisms combined with decentralized key management. This approach introduces a layer of decentralized trust by integrating TEE attestations directly into the blockchain, or a Data Availability alternative, ensuring that the integrity and state of the Trusted Execution Environment (TEE) can be verified independently and transparently by all network participants. This method draws inspiration from prior work, such as that of Flashbots~\cite{flashbots2024}, which has successfully implemented similar decentralized verification and key management techniques.

In contrast to User Attestation, where the user directly verifies the TEE instance via a local application and aTLS protocol, On-Chain Attestation relies on blockchain-based mechanisms to validate the TEE. Here, the TEE and user do not need to engage in direct trust-building steps at the beginning of a session. Instead, both TEE and the user encrypt their communications using asymmetric keys. The TEE’s attestation report, including its public key and proof of its state, is published on-chain, allowing any user to verify the integrity of the TEE through the blockchain before establishing a secure connection.

\paragraph{Decentralized Key Management.} A critical aspect of On-Chain Attestation is the use of decentralized key management, which ensures that cryptographic keys used by the TEE are securely handled and cannot be compromised by a single entity. When a TEE instance registers itself on the network, it generates a public-private key pair, where the public key is posted on-chain and associated with its attestation report. This decentralized key management approach removes the need for a centralized authority to manage keys, reducing the risk of single points of failure or trust. In addition, it improves transparency, since all participants in the network can independently verify the TEE’s credentials and encryption keys.

\paragraph{Asymmetric Encryption for Secure Communication.} Unlike User Attestation, where a direct secure connection is established using aTLS, On-Chain Attestation requires the use of asymmetric encryption for message exchanges between the user and the TEE. Once the user verifies the TEE’s public key and attestation report on-chain, they can encrypt their requests using the TEE’s public key, ensuring that only the TEE instance with the corresponding private key can decrypt and process the data. Similarly, the TEE encrypts its responses using the user’s public key, ensuring confidentiality in both directions of communication.

\paragraph{Advantages of On-Chain Attestation.} One of the key benefits of On-Chain Attestation is that it does not require specialized software to be installed on the devices of the users. Additionally, it further provides a more decentralized and transparent method for verifying TEE instances. Because attestation reports and public keys are stored on-chain, the verification process is open and trustless: Any participant can audit the information without relying on external parties. Additionally, this approach can scale more effectively in larger networks, as the verification process is transferred to the blockchain, eliminating the need for individual users to conduct direct attestations. This reduces the complexity for end users while maintaining strong security guarantees.

\paragraph{Challenges and Trade-offs.} However, On-Chain Attestation also presents challenges. The use of asymmetric encryption introduces performance overhead, as encrypting and decrypting large messages can be computationally intensive. Additionally, publishing attestation reports and public keys on-chain can lead to latency, depending on the speed and scalability of the underlying blockchain. Moreover, as this approach relies on blockchain transparency, care must be taken to avoid exposing sensitive metadata that could compromise privacy. Finally, decentralized key management, while reducing centralized trust, introduces complexity in securely handling key revocation and updating mechanisms.

% Alternatively, we can use on-chain verification combined with a decentralized key management.
% Similar approach has been used by preliminary work~\cite{flashbots2024}.
% Unlike User Attestation, the user and the TEE may need to encrypt messages using asymmetric keys.
% [@DLee] I think we need to be more precise what we mean by start here. Is the start, as in when the node registers for the first time, or when a connection user/node is established ?

\subsection{Node Registration}
When a node initiates a Trusted Execution Environment (TEE) instance, each TEE generates an asymmetric key pair—a public and private key—essential for secure communication and encryption. Following this, the TEE initiates a remote attestation process, binding its public key to the attestation report, which includes critical details about the TEE's state and its hardware configuration. This attestation process serves as a proof that the TEE instance is operating in a secure, trusted environment, isolated from any potentially malicious software or hardware interference.

The attestation report is then submitted for verification by an on-chain verification service. A notable example of such a service is Automata~\cite{automata}, which specializes in providing decentralized, trustless verification of attestation reports. Automata, or similar services, evaluates the integrity of the TEE’s attestation report, ensuring that it has been generated by legitimate, secure hardware, and that the TEE instance adheres to the specified security guarantees. This verification step is critical in maintaining trust across the decentralized network, as it prevents compromised or falsified TEE instances from participating.

Once the attestation report is successfully verified, the contract governing the decentralized AI network records the TEE’s public key on the blockchain. By storing the public key on-chain, the network ensures that all participants—whether they are users, other nodes, or external validators—can access and verify the TEE’s public key without relying on a centralized authority. This decentralized storage of public keys enables trustless encryption of communication between users and the TEE, ensuring that only the verified TEE instance can decrypt and process sensitive data.

The registration of the public key on-chain also facilitates the wider ecosystem's interactions with the TEE. For example, when a user wants to send a request to the TEE, they can fetch the public key from the blockchain and use it to encrypt their data, confident that only the corresponding TEE instance, with its associated private key, can decrypt and access the information. This cryptographic assurance protects both the user's data and the integrity of the AI computation being executed within the TEE.

In addition to recording the public key, the blockchain entry also serves as a public record of the TEE’s verified attestation status. This transparency allows anyone on the network to audit the TEE’s state and verify that the node is operating within the prescribed security parameters. By leveraging on-chain verification and decentralized key management, this node registration process ensures both security and trust in the decentralized AI network, mitigating risks such as unauthorized access, malicious computation, or tampering with sensitive workloads.

Finally, this on-chain record also facilitates future updates or key revocations. If a TEE instance becomes compromised or needs to be reinitialized, the network can easily revoke its public key, ensuring that compromised keys cannot be used to decrypt data or interact with the system. This flexibility enhances the security of the entire decentralized AI infrastructure, allowing for robust, trustless management of TEE instances across a distributed network.

% \paragraph{Node Registration.}
% Whenever a node starts TEE instances, each TEE instance generates a asymmetric key pair, and initiates a remote attestation with the public key bound to the attestation report.
% The attestation report is verified by on-chain verification service, such as Automata~\cite{automata}.
% Once the report is verified, the contract puts the public key on the blockchain.

\subsection{User Requests}

In the On-Chain Attestation model, users interact with the decentralized AI network via their web browser, establishing communication through the contract on the blockchain. When a user submits a request, the contract assigns a verified TEE instance to handle the task. This assignment is based on the attestation reports stored on-chain, ensuring that the TEE instance is trustworthy and operates within a secure, verified environment.

Once a TEE instance is assigned, the user is responsible for retrieving the TEE’s public key, which has been previously verified and published on the blockchain during the node registration process. This public key plays a crucial role in ensuring that the communication between the user and the TEE remains secure. The user must encrypt their input data—whether it's a simple text query or a more complex input like a large dataset or multimedia file (e.g., images or videos)—using this public key. The encrypted data is then posted back to the contract, ensuring that only the designated TEE instance, possessing the corresponding private key, can decrypt and process the user's input.

For use cases involving large files, such as image or video data, direct on-chain storage can be inefficient and expensive due to the limitations of blockchain networks. As a result, Atoma will leverage an external decentralized storage solutions, allowing for the scalable, cost-effective storage and retrieval of large data. In this scenario, the user would upload their large input files to the decentralized storage platform, and instead of posting the entire file on-chain, they would include a reference (e.g., a hash or URL) to the file location in their encrypted request.

The integration of decentralized storage ensures that even large inputs can be processed efficiently within the network, while maintaining the privacy and security guarantees of the On-Chain Attestation model. By encrypting only the reference to the file stored on decentralized storage, the user minimizes the amount of data handled directly by the blockchain, thereby optimizing the performance of the system without compromising security.

Once the TEE instance receives the user’s encrypted input and file reference, it can fetch the necessary data from the decentralized storage, decrypt the input using its private key, and begin processing the request securely. The user can rest assured that their sensitive data, whether small or large, is protected throughout the entire workflow—both while stored off-chain and while being processed on-chain.

Moreover, this architecture benefits from the transparency and auditability of the blockchain. Users can track the status of their request, verify that the assigned TEE instance is legitimate, and ensure that their input data remains secure at every step. This seamless integration of on-chain attestation, decentralized storage, and cryptographic encryption provides a powerful framework for privacy-preserving AI computations in decentralized environments.

% \paragraph{User Request.}
% In On-Chain Attestation model, the user can interact with the contract via their web browser.
% The contract will assign a TEE instance, and the user is responsible for pulling the public key of the TEE,
% encrypting their input, and posting it back to the contract.
% Since user input may contain some large files (e.g., image or video), the nodes may need to support fetching inputs via decentralized storage~\cite{ipfs,walrus}.

In the TEE Co-processor Execution model, each Trusted Execution Environment (TEE) instance operates as a dedicated, secure co-processor that actively monitors the blockchain for incoming tasks. The TEE instance continuously polls the blockchain until it detects that a job has been scheduled and assigned to it. This decentralized coordination ensures that tasks are dynamically distributed to verified TEE instances, enabling seamless execution of privacy-sensitive AI workloads.

Once a request is assigned to the TEE instance, the first step is for the instance to fetch the encrypted input associated with the job. This input may be stored on-chain or, in cases where the input is large (such as an image, video, or dataset), the TEE will retrieve a reference to the input from a decentralized storage solution. The TEE instance then decrypts the input using its private key, which was securely generated and managed during the node registration process. This ensures that only the specific TEE instance assigned to the task can access the decrypted input, maintaining the integrity and privacy of the user's data throughout the execution process.

Once the input is decrypted, the TEE performs the required computations to serve the request. This could involve running an AI model for inference, processing large datasets, or executing complex machine learning algorithms—all within the isolated and secure environment of the TEE, ensuring that no external software or hardware can interfere with or access the sensitive data being processed.

Upon completing the task, the TEE securely prepares the output. If the output is small, such as a prediction or a summary, it is encrypted and posted directly back to the blockchain, where it can be retrieved by the user or further processed by other nodes. For larger outputs—such as processed videos, datasets, or AI-generated media—the TEE may use decentralized storage to store the results. In this case, the TEE will upload the output to decentralized storage and post an encrypted reference (such as a hash or URL) back to the user, or to the blockchain directly. This reference allows the user to securely retrieve the output from the decentralized storage while ensuring that the data remains encrypted and accessible only to authorized parties.

The TEE's role as a co-processor ensures that all stages of the job—input retrieval, decryption, computation, and output handling—are conducted in a secure, isolated environment, minimizing the risk of unauthorized access or data leakage. The entire workflow benefits from the trustless nature of blockchain, where all interactions are recorded transparently and verifiably. Users can audit the execution process, verify that their input was processed by a legitimate TEE instance, and ensure that the output they receive is authentic and has not been tampered with.

Furthermore, by offloading the storage of large inputs and outputs to decentralized storage solutions, the TEE co-processor model achieves both scalability and efficiency. This approach allows the network to handle complex, data-intensive tasks while maintaining the high security standards required for decentralized AI computations. Through this combination of secure co-processor execution, blockchain-based transparency, and decentralized storage, the Atoma network provides a robust framework for privacy-preserving, distributed AI workloads.

% \paragraph{TEE Co-processor Execution.}
% The TEE instance constantly pulls the blockchain until it sees the scheduled job.
% When a request is assigned to the instance, it fetches the input, decrypt it, and then serve the request.
% The output is posted back to the blockchain, and potentially using a decentralized storage if necessary.

\subsection{Support for Current Architectures}

Currently, Automata, does not provide native support for the verification of confidential virtual machines (VMs) such as Intel Trusted Domain Extensions (TDX) or AMD Secure Encrypted Virtualization (SEV). These technologies are crucial for running secure, hardware-enforced environments within decentralized AI networks that ensure both the integrity and confidentiality of virtual machines.

To address this gap, it may be necessary to develop a custom verification solution specifically designed to support confidential VMs like Intel TDX and AMD SEV within the On-Chain Attestation framework. Building such a solution would involve implementing mechanisms that can verify the attestation reports generated by these confidential VMs, ensuring that they meet the same high security and integrity standards required for participation in the decentralized AI network.
Developing this custom verification layer involves several key steps:
\paragraph{}
\textbf{Attestation Integration for Confidential VMs:} First, we would need to integrate the attestation processes of Intel TDX and AMD SEV into the verification workflow. These confidential VMs generate their own unique attestation reports that provide cryptographic proof of the VM’s state, the hardware it runs on, and the isolation guarantees enforced by the hardware. By capturing these reports, we can ensure that each VM operates in a secure environment, isolated from other processes and potential threats.
\paragraph{}
\textbf{On-Chain Verification Logic:} Next, we would need to design and implement the on-chain logic for verifying the attestation reports generated by these confidential VMs. This could involve writing smart contracts capable of parsing and validating the cryptographic proofs included in the reports, ensuring that the VMs have not been tampered with and are running trusted, verified code. This on-chain logic would mirror the existing capabilities of Automata but would be extended to support these specialized hardware-based VM environments.
\paragraph{}
\textbf{Committee voting attestation verification:} An alternative to the previous suggestion, is to store TEE attestation in a decentralized storage solution, and allow a decentralized committee of nodes to check the validity of the posted attestations. Each committee participant can sign the result of attestation verification, and the final aggregate signature can be submitted to the blockchain, as a \emph{signed committee verification agreement}.
\paragraph{}
\textbf{Decentralized Key Management:} Once the attestation reports are verified, the public keys of the VMs could be registered on-chain, similar to the process for standard TEE instances. This decentralized key management would ensure that all participants in the network can trust the cryptographic credentials of the confidential VMs without relying on a central authority, aligning with the principles of decentralization and trustless verification.
\paragraph{}
\textbf{Scalability and Efficiency:} A critical consideration in developing this custom solution is ensuring that the verification process remains scalable and efficient. Attestation reports for confidential VMs can be complex and may require substantial computational resources to verify. The custom verification mechanism must be optimized to handle these reports without introducing significant latency or overhead to the network, particularly as the number of confidential VMs participating in the system grows.

% \subsubsection{Discussion}
% Automata currently does not provide verification on confidential VMs (e.g., Intel TDX or AMD SEV).
% Thus, we may need to build it by ourselves if we want to choose On-Chain Attestation model.

\subsection{Other Design Details}

\paragraph{Attestation Policy.}
% [@DLee] I think we should refer this is done upon node registration ? Do we know how large this combined data is ? I presume a few kBs ?
For attestation, the information about the expected TEE software stack and certificates should be kept up-to-date in the blockchain.
The policy consists of three parts: vendor policy, Atoma policy, and community policy.
The \textit{vendor policy} includes all information about accepted measurement of vendor-provided components: hardware, firmware, VMM (e.g., TDX SEAM), and drivers, and is governed by the hardware vendor.
The \textit{Atoma policy} includes accepted measurements of Atoma components: VM image for the execution node containing software distribution.
The up-to-date policy is maintained via predefined governing policy.
For example, each node may fetch the latest release of Atoma software stack from GitHub, and obtain the measurement.
Finally, the \textit{community policy} is an extra policy that can override the existing policies.
For instance, the community can agree to reject a certain measurement when they find a vulnerability.
We expect the combined policy to be a few tens of KBs assuming there will be less than a thousand measurements at any time.

\paragraph{Initial State Integrity.}

The attestation of TEE can verify the initial state of a system.
% [@DLee] Can we mention other cases ? Or are we sure we will only rely on confidential VMs (if I am not mistaken, Nvidia's requires confidential VMs for CPU/GPU communication, right?)
In confidential VMs such as Intel TDX or AMD SEV-SNP, it's the initial state of the virtual machine.
% [@DLee] Just to be sure, this is the attestion referred above correct ? (Root CA + GitHub, etc).
VMs are essentially a large state machine, where its state is exponential.
Thus, the initial state of the VM doesn't really tell much about the integrity of the software state when the user starts interacting with the TEE via a session.

We could leverage a few approaches to provide the initial state guarantee for each session in Atoma.
% [@DLee] I think it would be best to briefly explain what a confidential container is here.
First, we can rely on the confidential container technologies that are built on top of the CVMs.
Confidential containers aim to support attestation in container granularity, but rely on the software-based inter-container isolation.
By using confidential containers, we can bind each session to a fresh container, providing the confidence that the application and its environment were not compromised.
In addition, we can have the node rotate the CVM periodically such that it will reset the TEE's state to the benign state.
This also makes sure that any keys generated inside the TEEs are rotated (e.g., the asymmetric key of On-Chain Attestation model).

% [@DLee] I think it makes sense to separate the risk analysis below from the architectural design (two separate sections). I also suggest that the organization of the architectural section follows a more sequential flow, that is: 1) we mention each step involved at node registration (attestation generation), 2) how we can provide the guarantees for the nodes attestations 3) node/user/contract communication, etc.

% [@DLee] Can we add more information about how the node can setup their software/hardware at start to make sure they are running confidential compute ? We can refer to the hardware vendors guide for TEE initialization, etc.

\section{Security of the Design}

TEEs heavily rely on the design and implementation of the underlying hardware, thus may have unknown vulnerabilities that could potentially break the security guarantees.
In this section, we analyze risks of various attacks in the context of Atoma's decentralized AI,
and why we think TEE can provide privacy.

\subsubsection{General TEE Threat Model}

In general, TEEs provide robust security protections for applications by defending against several types of attacks:

\begin{itemize}
    \item \textbf{Physical Memory Access.} Attackers with physical access to hardware can potentially exploit vulnerabilities to read data from the main memory (i.e., DRAM) through techniques like the "cold-boot attack," which involves freezing the memory and retrieving active data. TEEs mitigate this threat by encrypting the contents of the main memory using hardware-protected keys that are inaccessible to software or external agents. This ensures that even if an attacker gains physical access to the hardware, they cannot read the encrypted memory data. However, it is important to note that TEEs typically do not encrypt GPU memory, as modern GPUs use 3D-stacked High Bandwidth Memory (HBM) technology, which makes it nearly impossible to physically extract data without destroying the chip itself.
    \item \textbf{Privileged Software.} Machine owners or attackers with administrative privileges could attempt to compromise a system by installing malicious software, modifying firmware, tampering with the operating system, or using root-level permissions to introduce bogus drivers. TEEs are designed to defend against such privileged adversaries by isolating the application from the underlying operating system and software stack, ensuring that even if the machine's firmware or OS is compromised, the application running within the TEE remains secure. TEEs achieve this level of protection by relying on cryptographic assurances enforced by the hardware, preventing unauthorized access or tampering.

\end{itemize}

On the other hand, TEEs rely on a few assumptions:

\begin{itemize}
    \item \textbf{Application Has No Security Issue.} TEEs offer robust protection against physical and privileged attacks, but they cannot address vulnerabilities or flaws inherent in the application itself. If the application has security weaknesses—such as poorly designed code, unpatched vulnerabilities, or intentional backdoors—TEEs are unable to rectify these issues. The responsibility for securing the application’s logic, architecture, and codebase lies with the software developers. TEEs only protect the execution environment, not the internal integrity of the software. At Atoma, we are committed to providing a thoroughly audited, secure software stack, designed to minimize potential risks and ensure that applications built on our platform are as secure as possible.
    \item \textbf{Hardware Has No Vulnerability.} Another key assumption is that the hardware running the TEE is free from vulnerabilities or design flaws. While hardware is generally more resilient to attacks compared to software—due to its smaller attack surface and the rigorous verification processes it undergoes—there have been documented cases where hardware vulnerabilities were exploited, as highlighted by research on attacks such as WeSee~\cite{wesee}, Heckler, and Foreshadow~\cite{foreshadow}. These studies demonstrate that even hardware is not immune to security flaws. Nevertheless, hardware vulnerabilities are less frequent than software vulnerabilities, and TEEs still provide a significant layer of defense in most cases. In Section~\ref{sec:attack-tee}, we will discuss why we believe that, despite these potential hardware vulnerabilities, TEEs remain a viable and valuable tool for enhancing security in the Atoma network.

\end{itemize}

Side channel attacks represent a method by which attackers attempt to extract sensitive information by analyzing indirect data from various parts of a system. Rather than directly accessing private data, attackers gather clues from observable system behaviors such as execution time, memory access patterns, power consumption, or even physical characteristics like heat dissipation. By correlating these side-channel observations, attackers may infer secrets from running applications. Although Trusted Execution Environments (TEEs) provide robust protection against direct data access, they are not fully immune to side channel attacks.

Recent research into attacks on Intel’s SGX (Software Guard Extensions) has shown that even TEEs can be vulnerable to certain side channel techniques. These attacks exploit subtle information leaks that occur as programs execute, such as differences in the time it takes to perform specific computations or variations in memory access patterns. Theoretically, it is impossible to completely eliminate all side channels, as some form of information leakage will always be present in any computational system. However, many side channel attacks are constrained by low bandwidth and high cost, making it difficult for attackers to extract useful information quickly or affordably enough to mount practical attacks.

Among the most feasible side channel attacks are those that exploit memory access patterns, such as cache side-channel attacks. These attacks are particularly effective in systems based on the \textit{Von Neumann} architecture, where applications use shared memory resources. When a program executes, it may access memory in patterns that depend on the underlying data being processed. In scenarios where an application contains data-dependent conditional branches, an attacker observing the memory access patterns may be able to deduce information about the data being processed by the application. This is especially concerning in sensitive applications where even small information leaks could reveal critical data.

In the context of Atoma, however, the risk of side channel attacks is significantly reduced due to the nature of the workloads being processed. The primary tasks performed by nodes in Atoma involve machine learning inference and training, which rely heavily on Basic Linear Algebra Subprograms (BLAS) libraries. These programs involve extensive matrix and vector operations, which are typically executed in a highly parallelized manner. Crucially, these computations do not generally involve data-dependent branches—meaning that the memory access patterns remain consistent regardless of the specific data being processed. This lack of data-dependent control flow significantly reduces the attack surface for timing and access-pattern-based side channels.

To date, there are no well-documented side channel attacks that target matrix or vector-based computations, such as those commonly found in machine learning workloads. As a result, Atoma’s reliance on highly parallelized numerical operations, which are largely independent of the specific data being processed, provides a natural resilience against the most common types of side channel attacks.

While it is important to acknowledge that no system can ever be completely immune to all forms of side channels, Atoma’s architecture minimizes the practicality of such attacks. The specific nature of the computations involved, combined with the robust protections provided by TEEs, makes it highly unlikely that Atoma will be vulnerable to general side channel attacks. Nonetheless, ongoing monitoring and research into emerging side channel techniques will be critical to maintaining the network’s security in the long term.

\subsubsection{Attacks on TEE}
\label{sec:attack-tee}

As mentioned earlier, the security guarantees provided by Trusted Execution Environments (TEEs) are contingent on the correct implementation of the underlying hardware, as well as the supporting manufacturer software, such as microcode and firmware. However, no system is entirely immune to vulnerabilities, and it is theoretically impossible to ensure that any complex system is completely free of bugs. Over the years, several studies have demonstrated sophisticated attacks capable of partially compromising the security properties of TEEs~\cite{wesee,heckler}.

Fortunately, the vast majority of these vulnerabilities are identified and patched swiftly, often before they become publicly known. Many of these attacks are discovered by ethical researchers who work with hardware vendors to address the issue before it is disclosed to the public. This proactive approach helps mitigate potential threats early in the vulnerability lifecycle. Additionally, even in the case of a zero-day exploit—an attack that takes advantage of an undiscovered vulnerability—the decentralized AI network can implement contingency measures, such as temporarily disabling the privacy-preserving mode of TEEs through a community-driven policy, ensuring the system remains secure while patches are developed and deployed.

While there remains the possibility of an unethical actor discovering a TEE vulnerability and exploiting it to compromise user privacy, we argue that such scenarios are highly improbable. There are several reasons why the likelihood of such an attack is low:
\paragraph{}
\textbf{High Cost of Attack:} Conducting a successful attack on a TEE typically requires considerable expertise and resources, often involving black-box analysis of proprietary hardware. These attacks are extremely complex and can take years to succeed. Even then, there is no guarantee that a vulnerability exploited in older hardware will remain effective in the latest generation of TEE implementations, as manufacturers regularly improve security in newer designs.
\paragraph{}
\textbf{Limited Financial Incentive:} The decentralized nature of the Atoma network, which uses randomized node selection, makes targeted attacks nearly impossible. An attacker cannot choose a specific node to exploit, meaning the best possible outcome for the attacker is gaining access to the data of random users. Without a clear financial incentive or the ability to launch a focused attack on high-value targets, the effort required for such an exploit far outweighs any potential gain.
\paragraph{}
\textbf{Combining Multiple TEE Technologies:} To further enhance security, Atoma can implement a layered defense by combining more than one TEE technology. For instance, users could opt to double-encrypt their input using two different TEE platforms—such as Intel TDX and AMD SEV—before it is processed. This dual encryption strategy means that in order to compromise user privacy, an attacker would need to break both TEE implementations simultaneously. Since each TEE platform is based on distinct hardware architectures and security models, successfully exploiting both at the same time would raise the complexity and cost of the attack exponentially, making it highly unlikely.
\paragraph{}
By leveraging multiple TEE technologies and employing layered security strategies, the Atoma network significantly raises the difficulty and cost of any potential attack, further safeguarding user privacy. While no system can ever be guaranteed to be completely invulnerable, the combination of proactive patching, decentralized node selection, and multi-layered TEE defenses makes attacks on Atoma’s privacy-preserving features highly impractical. This comprehensive approach allows the network to continue providing strong privacy guarantees in decentralized AI environments, even in the face of evolving security challenges.

\section{Future Work}

\paragraph{TEE for verifiability.}
We are investigating the possibility of using TEEs for verifiability in Atoma network.
We believe that the attestation capability of TEEs can compliment our Sampling Consensus by raising the bar of attacking the integrity.
Simply put, TEEs can provide the proof of execution for all requests (including non-private requests).

\paragraph{Model Privacy}
\label{sec:future-model-privacy}
Currently, Atoma only aims to provide open source models.
However, with TEEs, we can also provide model privacy, unlocking a new opportunity to serve private models.
In this case, we may need an extra policy that is determined by the model owner, such that they can decide which measurement they would like to reject.
Also, we will need to address potential attacks on TEEs as the attacks now can have financial incentives.
Although targeted attacks are still not possible, the advantage of the attack may exceed the cost of the attack depending on the value of the model.
We have a few ideas to mitigate such attacks, and are actively exploring options.

\bibliographystyle{plain}
\bibliography{references}

\end{document}